\documentstyle[proceedings,numreferences,epsf]{crckapb}
\newcommand{\dfrac}[2]{\displaystyle\frac{#1}{#2}}
\newcommand{\mathrm}[1]{{\rm #1}}
\newcommand{\mathbf}[1]{{\bf #1}}
\newcommand{\text}[1]{\mbox{#1}}

\begin{opening}
\title{Solar wind acceleration by the dissipation of Alfv\'en waves}
\author{A.~A.~Galeev}
\author{A.~M.~Sadovski}
\institute{Space Research Institute of Russian Academy
 of Sciences\\
117810, Profsouznaya 84/32, GSP-7, Moskow, Russia}
\end{opening}
\begin{document}
\begin{abstract}
Axford and McKenzie [1992] suggested that the energy released
 in impulsive reconnection events generates 
high frequency Alfv\'en waves. The kinetic equation for
 spectral energy density of waves is derived in the 
random phase approximation. Solving this equation we find
 the wave spectrum with the power law ``-1'' in the
 low frequency range which is matched to the spectrum
 above the spectral brake with the power low ``-1.6.''
The heating rate of solar wind protons due to the dissipation
 of Alfv\'en waves is obtained.
\end{abstract}
\section{Introduction}

The solar wind which is 
 formed by the gas-dynamic expansion of solar corona
 into the interplanetary space can be divided into 
two states: the fast
($>700$~km/s)
and slow
($\approx 400$~km/s)
streams. The fast solar wind originates in the polar coronal
 holes,  regions of the solar corona with relatively low
 electron temperature
$T\approx 10^6$~K,
reduced plasma density 
$n=10^8$~cm${}^{-3}$
and magnetic field at the base of corona 
$B_0\approx 10$~G~\cite{Axford},~\cite{Parker}.
Variations of the velocity of this fast solar wind streams are
 insignificant 
(700--800~km/s).
The slow solar wind is associated with transient openings of
 closed field regions in the corona. The streams of slow wind
 are limited by the range 
$\pm 13^{\circ}$
near the equator. The transition zone between this regions
 has the width 
$\pm 13^{\circ }-\pm20^{\circ }$
on the north and south latitudes in the solar
 minimum~\cite{Schwenn}. The radiation balance analysis in the
 near equatorial coronal holes~\cite{Withbroe} shows that to
 accelerate solar wind it is necessary to dissipate the energy
 flux of radiationless origin 
$\approx(5\pm1)\cdot 10^5$~erg/cm${}^2\cdot$s
within the distance of one-two solar radii above the Sun's
 surface in order to maintain the observed temperature 
$1,5\cdot 10^5$~K
of slowly rising expanding gas.

Axford and McKenzie~\cite{Axford} suggested that the energy
 source, necessary to accelerate fast solar wind, is in the
 regions of strong magnetic field that define the boundaries
 of chromospheric supergranules.
If the magnetic field is not strictly unipolar in this regions,
 for example, the closed loops of magnetic field lines are
 existing at the base of the corona, then the necessary energy
 is released in the processes of impulsive reconnection of
 magnetic field lines on the characteristic spatial scale
 of network activity
$l=100$~km.
Such processes of impulsive reconnection are accompanied by the
 generation of Alfv\'en and fast magnetosonic waves.
Using the parameters of plasma at the coronal base described
 above we can find the Alfv\'en wave velocity
$V_A= B/\sqrt{4\pi nm_i} = 2\cdot10^8$~cm/s,
the ion's thermal velocity
$V_{Ti}=2\cdot10^7$~cm/s,
and the characteristic time of magnetic field lines
 reconnection~\cite{Petschek}:
\begin{equation}
\tau_R =
 \frac{\pi V_A}{4l\ln \mathrm {Re_m}}
 = \frac{\pi \cdot 2 \cdot 10^8 \quad \mbox{cm/s}}
      {4 \cdot 10^7 \quad \mbox{cm}\cdot \ln 3.5 \cdot 10^{11}}
 \sim 0.6 \quad \mbox{s},
\label{1}
\end{equation}
where:
$\mathrm {Re_m} = 4 \pi n e^2 V_A l / m_e \nu_{ei} c^2
                      = 3.5 \cdot 10^{11}$
is the magnetic Reinold's number,
$\nu_{ei} = 20 n / (T^\circ)^{3/2} = 2.3$~s${}^{-1}$
is the collision frequency.
It is evident that the characteristic period of Alfv\'en waves
 is of the order of obtained reconnection time.
The total energy flux necessary to accelerate  fast solar wind
 is estimated as 
$8\cdot10^5$~erg/cm${}^2\cdot$s~\cite{Axford} 
and as in the case of the near equatorial coronal hole the major
 part of the energy of generated waves should be dissipated
 within one-two solar radii above the solar surface.

As the dominant mechanism of dissipation of Alfv\'en waves
 generated by the processes of impulsive reconnection of
 magnetic field lines, we will consider the induced scattering
 of Alfv\'en waves by plasma ions~\cite{LeeVolk}.
For the sake of simplicity as in the paper~\cite{LeeVolk} we
 will limit ourselves by the case of the circularly polarized
 Alfv\'en waves propagating along the radial magnetic field 
$\mathbf B_0 = B_0(r) \mathbf z$
from the polar coronal holes with the wave vector 
$\mathbf k = (0,0,k)$,
where 
$r$
is the radial distance from the center of the Sun.
With the help of the above cited parameters of plasma and
 magnetic field at the base of the corona, we fix the following
 ordering of the ion cyclotron
$\omega_{ci}$,
Alfv\'en 
$\omega_k$
and ion Doppler
$kv_{Ti}$
frequencies:
\begin{equation}
\omega_{ci}\gg\omega_k\gg kV_{Ti},
\label{2}
\end{equation}
where
$\omega_{ci}=eB_0(r)/m_ic$,
$\omega_k=kV_A$.
Under this conditions plasma is magnetized and the resonant
 condition describing the interaction of Alfv\'en waves with
 the different polarization propagating in the opposite
 directions along the magnetic field lines can be written as:
\begin{equation}
\omega_k + \omega_{k'} - \left(k+k'\right)v_{z}=0,
\label{3}
\end{equation}
where
$\omega_k =kV_A > 0$, $\omega_{k'} =-k'V_A < 0$, $k,k'>0$ 
and
$v_z$
is the velocity along the magnetic field.
Here in the process of induced scattering the quantum of
 higher frequency is exchanged for that of lower frequency.
As a result of this some fraction of wave energy is transferred
 to plasma ions.

In the next section we will derive the kinetic equation
 for the interacting waves in the random phase approximation.
 In this derivation we limit ourselves by the case
$T_i\gg T_e$~\cite{LivTsy},~\cite{Tsytovich},
that is valid for the solar corona.
Then in the section 3 we will describe the evolution of
 the Alfv\'en waves spectrum in the processes of their
 propagation and show that the wave energy flux necessary 
to accelerate the fast solar wind can be dissipated at the
 heights of 1-2 solar radii above the solar surface that is
 required by the Withbroe~\cite{Withbroe} analysis. 

\section{Derivation of kinetic equation in the
 random phase approximation}

The kinetic equation for the particle distribution function 
$f_j$
in the absence of collisions takes the form~\cite{Galeev},~\cite{Lyons}:
\begin{equation}
\begin{array}{l}
\displaystyle
\left( \frac{\partial}{\partial t} + v_z 
\frac{\partial}{\partial z} \mp \omega_{cj} 
\frac{\partial}{\partial \theta} \right) \cdot f_j =
\frac{ie_j}{2m_jc}B^\pm_k e^{\mp i\theta}
\left[ \left( \frac{\omega_k}{k}-v_z \right)
\frac{\partial}{\partial v_\perp}+
\right.\\
\displaystyle
\qquad\qquad\left.
+ \frac{\partial}{\partial v_z} v_\perp
\mp i \left( \frac{\omega_k}{k}-v_z \right)
\frac{\partial}{v_\perp\partial\theta} \right]
\cdot f_j\exp\left[-i\left(\omega_kt-kz\right)\right],
\label{4}
\end{array}
\end{equation}
where:
$v_\perp$ and 
$\theta$ 
are the velocity and the asimuthal angle of the cyclotron
 rotation of particles, 
$B^+_k$ and
$B^-_k$ 
are the waves amplitudes with the left and the right
 polarizations propagating away from Sun and towards the Sun respectively.

Following  perturbation theory we solve the
 kinetic equation (\ref{4}) by finding the distribution function
 of plasma particles as power series in the wave amplitudes:
\begin{equation}
f_j=f_{j0}+f_{j1} +f_{j2} +f_{j3},
\label{5}
\end{equation}
where 
$f_{j0}$ 
is the isotropic maxwellian distribution of particles of $j$
 species in the zero approximation and 
$f_{jn}$ 
is the particle distribution function on the $n$-th step of iteration.
Using the iteration procedure for solving the kinetic equation
 (\ref{4}) we obtain the corresponding Fourier's components of
 the particle distribution functions for the first two steps of iteration:
\begin{equation}
f_{j1}^+ = - \frac{e_j\omega_k}{2m_jkc}
\frac{\partial f_{j0}/\partial v_\perp}
{\omega_k-kv_z-\omega_{cj}} B^+_k e^{-i\theta} \exp \left[ -i 
(\omega_k t - kz) \right],
\label{6}
\end{equation}
\begin{equation}
f_{j1}^- = \frac{e_j\omega_{k'}}{2m_jk'c}
\frac{\partial f_{j0}/\partial v_\perp}
{\omega_{k'}-k'v_z+\omega_{cj}} B^-_{k'} e^{i\theta} \exp 
\left[-i (\omega_{k'} t - k'z) \right],
\label{7}
\end{equation}
\begin{equation}
\begin{array}{l}
\displaystyle
f_{j2} = - \dfrac{\omega_{cj}}{2} V_A \left[ 1 + \dfrac{\left( 
k+k' \right) V_A}{2 \omega_{cj}} + \dfrac{\left( k^2+k^{\prime 2} 
\right) V_A^2}{2 \omega_{cj}^2}+ \dfrac{\left( k-k' \right) 
v_z}{2 \omega_{cj}} \right.\\
\qquad\qquad\left.+ \dfrac{\left( k^{\prime 2}-k^2 \right) V_A v_z}{
\omega_{cj}^2} \right] \dfrac{\partial f_{j0}/\partial v_z}
{\omega_k + \omega_{k'} - \left( k+k' \right) v_z + i0} \\
\qquad\qquad\times
\dfrac{B^+_k B^-_{k'}}{B^2_0} \exp \left[-i \left( \omega_k + 
\omega_{k'} \right) t + i \left( k+k' \right) z \right].
\label{8}
\end{array}
\end{equation}
Here we have neglected the small contribution related to the
 differentiation of nonresonant denominator on
$v_z$ in the equation (\ref{8}).

Perturbation of the charge density related to the particle
 distribution function of the second order 
$f_{j2}$
excites the longitudinal wave with the potential
$\phi_{k+k'}$,
that can be found from the Poisson equation:
\begin{equation}
\left( k+k' \right)^2 \phi_{k+k'} \epsilon_{k+k'} = 4\pi
 \sum\limits_{j}^{} e_j\int\,d^3 {\bf v} f_{j2},
\label{9}
\end{equation}
where
 $\epsilon_{k+k'}$
 is the dielectric permeability of the plasma:
\begin{equation}
\epsilon_{k+k'} = 1 + \sum\limits_{j}^{} 
\frac{\omega_{pj}^2}{\left( k+k' \right)^2} \int\,d^3 {\bf v} 
\frac{\left( k+k' \right) \partial f_{j0}/\partial 
v_z}{\omega_k+\omega_{k'}-\left( k+k' \right)v_z + i0}.
\label{10}
\end{equation}
With the help of the operator of particle interaction with the
 longitudinal waves in a form:
\begin{equation}
f_{j\phi} = -\frac{e_j}{m_j} \frac{\left( k+k' \right) 
\phi_{k+k'}}
{\omega_k + \omega_{k'} - \left( k+k' \right)v_z+i0}
\frac{\partial f_{j0}}{\partial v_z}
e^{-i \left( \omega_{k}+\omega_{k'} \right)t +i \left( k+k' 
\right)z}
\label{11}
\end{equation}
we find on the next iteration step the particle distribution
 function of the third order for the waves propagating from Sun:
\begin{equation}
\begin{array}{l}
 f_{j3}^{-*} = -\dfrac{e_j}{2m_jc} \displaystyle\int\,
 \dfrac{dk'}{2\pi}
\dfrac{\left[ \displaystyle \left(
\dfrac{\omega_{k'}}{k'} - v_z \right)
\displaystyle \dfrac{\partial}{\partial v_\perp} +
v_\perp \displaystyle \dfrac{\partial}{\partial v_z} \right]}
{\omega_k - kv_z - \omega_{cj}+i0}
\left( f_{j2} + f_{j\phi} \right)\\
\qquad\qquad\qquad\times B^{-*}_{k'} e^{-i\theta}
\exp\left[i(\omega_{k'}t - k'z)\right].
\label{12}
\end{array}
\end{equation}

The kinetic equations for the waves propagating from Sun and
 towards the Sun respectively can be described in terms of
 Maxwell equations for the Fourier's components of the wave amplitudes:
\begin{equation}
B^\pm_k=\mp \frac{4\pi}{kc} \sum\limits^{}_j e_j
\int\, d{\bf v} v_\perp e^{\pm i\theta}
\left( f^\pm_{1j} + f^{\mp*}_{3j} \right),
\label{13}
\end{equation}
Here we take into account that the second order current,
 associated with the distribution functions
$f_{j2}$ and $f_{j\phi}$, does not  contribute
 to~(\ref{13}).

In the linear approximation the dispersion relation~(\ref{13})
 can be written as~\cite{David}:
\begin{equation}
\left\{ 1 - \sum\limits_j^{} \frac{\omega_{pj}^2
\omega}{2k^2 c^2} \int_{}^{}\,d^3 {\bf v} \frac{v_\perp
\partial f_{j0}/{\partial v_\perp} }
{\omega_k-kv_z-\omega_{cj}+i0}\right\}B^+_k = 
\epsilon^+_{k1} B^+_k =0.
\label{14}
\end{equation}
Taking the frequencies ordering (\ref{2}) we obtain from the
 equation  (\ref{14}) the well-known dispersion relation for the
 Alfv\'en waves:
\begin{equation}
\epsilon^+_{k1}\approx 1-
\frac{\omega^2_k}{k^2v^2_A}
=0
\label{15}
\end{equation}

In the order to obtain the time dependent evolution of Alfv\'en
 waves we multiply the equation  (\ref{14}) by
$B^{+*}_k(k,\tilde\omega)\exp[i(\tilde\omega -\omega)]$
and integrate it by
$d\tilde\omega d\omega$.
Using the relation
$\epsilon^+_{k1}={\epsilon^+_{k1}}'+{i\epsilon^+_{k1}}''$,
we obtain:
\begin{equation}
\begin{array}{l}
\displaystyle
 \int\!\!\int\, d\tilde\omega d\omega \epsilon^+_{k1}
B^+_k(k,\omega)
 B^{+*}_k(k,\tilde\omega) e^{i(\tilde\omega  -
\omega)t}\\
\displaystyle\qquad\qquad\qquad
 = \frac i2
 \frac{\partial{\epsilon^{+}_{k1}}'(\omega_k)}{\partial\omega_k}
\frac{d|B^+_k|^2}{dt} +
 i{\epsilon^{+}_{k1}}''(\omega_k) |B^+_k|^2,
\end{array}
\label{16}
\end{equation}
where the time dependent wave amplitude is defined as:
\begin{equation}
B^+_k(t) = \int\,d\omega B^+_k(k,\omega) e^{-i(\omega  -
\omega_k)t}.
\label{17}
\end{equation}
Restoring the contributions of the nonlinear currents on the
 right-hand side of the equation (\ref{13}) we find the kinetic
 equation for the weakly interacting waves propagating from the Sun:
\begin{equation}
\begin{array}{l}
\displaystyle
 \frac{d| B^+_k |^2}{dt} = Im \sum\limits_{j}^{}
\frac{\omega_{pj}^2 \omega_k}{kc^2} \int\frac{dk'}{2\pi}
\int\,d^3 {\bf v} \frac{\omega_{k'}/k' - v_z}{\omega_k-
kv_z-\omega_{cj}+i0}\\
\displaystyle\qquad\qquad
\times\left( f^+_{j2} + f^+_{j\phi} \right) B^{+*}_k B^{-*}_{k'}
e^{i(\omega_k-\omega_{k'})t - i(k-k')z},
\end{array}
\label{18}
\end{equation}
To obtain the above equation in such form we integrate by
 parts the expressions for nonlinear currents to get rid of
 the derivatives over 
$v_z$ and $v_\perp$
and neglect small corrections related to the differentiation
 of the nonresonant denominator by
$v_z$.

Expanding the expressions for the nonresonant denominator and
 for the particle distribution functions 
$f_{j2}$ and $f_{j\phi}$ as a power series in 
$\omega_k/\omega_{cj}$ and
$\omega_{k'}/\omega_{cj}$,
$kv_z$ and $k'v_z$
to the second order we find the equation for waves propagating
 from the Sun in the form:

\begin{equation}
\begin{array}{l}
\displaystyle
\dfrac{d}{dt} |B^+_k|^2 = -Im \sum\limits_j
\dfrac{\omega_{pj}^2 V_A^2}{2c^2}
\int \dfrac{dk'}{2\pi} \int\,d^3 {\bf v}
\dfrac{\omega_k}{k} 
\left(
\dfrac{\omega_{k'}}{k'} - v_z
\right)
\left(
1+\dfrac{kV_A}{\omega_{cj}}- \dfrac{kv_z}{\omega_{cj}}
\right) \\
\times\left\{
      1+A_j - 
\displaystyle
      \dfrac{\sum\limits_{j'} \omega_{pj'}^2 \int\,d^3 {\bf v} 
      \left(
            1+A_{j'}
      \right)
      \dfrac{\partial f_{j'0}/\partial v_z}{\omega_{k}
     +\omega_{k'} - 
      \left(
            k+k'
      \right)
      v_z + i0}}{\sum\limits_{j'} \omega_{pj'}^2
      \int\,d^3 {\bf v} 
      \dfrac{\partial f_{j'0}/\partial v_z}{\omega_{k}
     +\omega_{k'} - 
      \left(
            k+k'
      \right)
      v_z + i0}}
\right\}\\
\displaystyle
\times\dfrac{\partial f_{j0}/\partial v_z}{\omega_{k}+\omega_{k'} - 
      \left(
            k+k'
      \right)
      v_z + i0}
\dfrac{|B^+_k|^2|B^-_{k'}|^2}{B_0^2},
\label{19}
\end{array}
\end{equation}
where
$$
A_j= \dfrac{\left(k+k'\right)V_A}{2\omega_{cj}}+
\dfrac{\left(k^2+k^{\prime 2}\right)V_A^2}{2\omega_{cj}^2}+
\dfrac{\left(k-k'\right)v_z}{2\omega_{cj}}+
\dfrac{\left(k^{\prime 2}-k^2\right)V_Av_z}{\omega_{cj}^2},
$$
and $A_{j'}$  and differ from $A_{j}$ only by the index $j'$.

Considering the limit
$T_e\ll T_i$
we obtain in the last term in figure brackets that the electron
 contribution to integrals dominates that for ions:
\begin{equation}
\begin{array}{l}
\displaystyle
\sum\limits_{j'} \int\,d^3 {\bf v}
\dfrac{\partial f_{j'0}/\partial v_z}{\omega_{k}+\omega_{k'} - 
      \left(
            k+k'
      \right)
      v_z + i0}\\
\displaystyle
\qquad\qquad\quad
\approx -\dfrac{\omega_{pe}^2}{k+k'} \int\,d^3 {\bf v}
\dfrac{\partial f_{j0}}{v_z\partial v_z} \gg 
\dfrac{\omega_{pi}^2}{\left( k+k'\right)v_{Ti}^2}.
\label{20}
\end{array}
\end{equation}
As the result the first term in the brackets cancels with the
 last one approximately equal 
$-1$.
Taking the integral by parts over 
$v_z$ and then over
$dk'$
we reduce equation ~(\ref{19}) to the equation obtained by 
Livshits and Tsytovich~\cite{LivTsy},~\cite{Tsytovich}:
\begin{equation}
\dfrac{d}{dt} |B^+_k|^2 =- \omega_k \dfrac{|B^+_k|^2 k}{B_0^2}
 \dfrac{\partial}{\partial k}
|B^-_k|^2 k.
\label{21}
\end{equation}
In the case of stationary expansion of plasma from the solar
 corona along approximately radial magnetic field lines this
 equation takes the form:
\begin{equation}
\left[
      u(r)+V_A(r)
\right]
\dfrac{\partial}{\partial r} |B^+_k|^2 = -k V_A
\dfrac{|B^+_k|^2 k}{B_0^2(r)}
\dfrac{\partial}{\partial k} |B^-_k|^2 k.
\label{22}
\end{equation}
\section{Evolution of Alfv\'en wave spectrum in the processes
 of their propagation}

Evolution of Alfv\'en wave amplitudes in the processes of their
 propagation into the interplanetary space we calculate
 theoretically assuming that (a) the interplanetary magnetic field 
$B_0(r)$ 
and the solar wind velocity
$u$
are radial and spherically symmetric, (b) the Alfv\'en waves
 are circularly polarized, (c) wave amplitudes are small
$|B^\pm_k(r)|^2 k \ll B_0^2(r)$.

Integrating the equation for the spectral energy density of
 waves\footnote{In the works~\cite{LivTsy},~\cite{Tsytovich},
 authors were interested in stationary spectra of Alfv\'en
 waves when the left-hand side of the equation has been zero,
 in the other words $\partial/\partial t=0$. That's why they
 didn't call any attention that the sign in the right-hand
 side of the equation should be mines not plus.}
\begin{equation}
\left[
      u(r)+V_A(r)
\right]
\dfrac{\partial}{\partial r} \dfrac{|B^+_k(r)|^2k}{4\pi}
 = -\kappa k^2 V_A(r)
\dfrac{|B^+_k(r)|^2 k}{B_0^2(r)}
\dfrac{\partial}{\partial k} \dfrac{|B^+_k(r)|^2k}{4\pi},
\label{+1}
\end{equation}
we find the general solution~\cite{Kamke}:
\begin{equation}
|B^+_k(r)|^2k = \Phi
\left[
     \kappa I \int\limits^r_{R_\odot} \dfrac{V_A(r)dr}{u+V_A(r)}
     +\dfrac{1}{k(R_\odot)} - \dfrac 1k
\right].
\label{+2}
\end{equation}
Here: 
$\Phi$ is an arbitrary function;
$V_A(r) = V_A(R_\odot)R_\odot/r$; 
$k(R_\odot) V_A(R_\odot)/2\pi=1.7$ Hz;
$R_\odot = 6.96 \cdot 10^{10}$ cm;
$\kappa = |B^-_k(r)|^2/|B^+_k(r)|^2$
is the ratio of the spectral energy density of Alfv\'en waves,
 propagating to the Sun and from the Sun respectively, that
 is the constant because both waves have the same spectral index 
$\alpha=-1$ 
and radial profiles. The ratio of the wave energy density to
 the energy of the magnetic field 
$I=|B^+_k(r)|^2k/B_0^2(r)$
happens to be the constant too due to the identical radial
 profiles of the energy density of Alfv\'en waves and the
 energy of magnetic field and very weak dependence of the
 wave energy density 
$|B^+_k(r)|^2k$ 
on wave vector 
$k$
due to the very small left-hand side of the equation~(\ref{+1}).
Numerical value for 
$I$ can be found by  equating the energy flux of Alfv\'en
 waves generated at the base of the corona to the energy flux
 of the radiationless nature with the value 
$8\cdot 10^5$~erg/cm${}^2\cdot$s,
necessary for the acceleration of the fast solar wind from
 polar coronal holes:
\begin{equation}
V_A(r)\dfrac{|B^+_k|^2k}{4\pi} = 
8\cdot 10^5 \text{{}~erg/cm${}^2\cdot$s}.
\label{+3}
\end{equation}
As the result we find that:
 $I=5\cdot 10^{-4}$.

In the process of wave propagation to the interplanetary space
 the position of the spectral break, separating spectral domains
 with the different slopes, is shifted to the low
 frequencies~\cite{Tu}. From the general solution~(\ref{+2})
 we construct the particular solution:
\begin{equation}
\dfrac{1}{k_{br}}= \kappa I \int\limits_{R_\odot}^r
 \dfrac{V_A(r)dr}{u+V_A(r)}
     +\dfrac{1}{k(R_\odot)},
\label{+4}
\end{equation}
We can find the exact position of the spectral break frequency
 as a function of radial distance by dividing both sides of
 this equation by the Alfv\'en wave velocity:
\begin{equation}
f_{br}(r) =
\left[
 0.6 +
2\pi \kappa I \int\limits^r_{R_\odot} \dfrac{V_A(r)dr}{u+V_A(r)}
\right]^{-1}\dfrac{ R_\odot }{r}.
\label{+5}
\end{equation}
Assuming that the acceleration of fast solar wind is terminated at
$r=3R_\odot$~\cite{13} we find that 
$f_{br}(3R_\odot)\approx 0.28$~Hz.
At this stage most of the heating has been deposited within
 $(1-2)R_\odot$ above the surface as required by
 Withbroe~\cite{Withbroe} analysis.
We can simplify this expression at the distance 
$r>10R_\odot$ 
assuming that the velocity of the fast solar wind
$u=7.5\cdot 10^7$~cm/s {}$\gg V_A(r)$.
As a result we reduce the expression~(\ref{+5}) to the form:
\begin{equation}
f_{br}(r) =
\left[
 0.6 +
\pi \kappa \ln\dfrac{r}{R_\odot}
\right]^{-1}\dfrac{R_\odot}{r}.
\label{+6}
\end{equation}
Taking the numerical value of the spectral break 
$6\cdot 10^{-2}$ Hz
at the distance
$r=60R_\odot $ ~\cite{Tu},
separating the domain of Alfv\'en wave turbulence with the
 spectral indexes 
$\alpha=-1$
and domain with
$\alpha=-1.6$
and equating it to the theoretical value obtained from
 the equation~(\ref{+6}) we find the ratio of the spectral
 energy density of Alfv\'en waves propagating to the Sun
 and from the Sun respectively that is equal to 
$\kappa=0.17$.
This means that the intensity of waves propagating to the
 Sun is six times lower than those for the waves propagating
 from the Sun.

Experimental observations show that above the spectral break
a wave spectrum with a slope near $-1.6$~\cite{Bav} is
established:
\begin{equation}
|B^+_k(r)|^2 = |B^+_{k_{br}}(r)|^2k_{br}^{1.6}/k^{1.6}.
\label{+7}
\end{equation}

The equation for the spectral energy density of waves in this 
case should inherit  the structure of the equation (\ref{+1})
with the continuos transition across the spectral break:
\begin{equation}
\begin{array}{l}
\left[
      u(r)+V_A(r)
\right]
\dfrac{\partial}{\partial r} \dfrac{|B^+_k(r)|^2k^{1.6}}{4\pi}\\
\qquad\qquad\quad = -\kappa k^{1.6} V_A(r)k_{br}^{-0.2}
\dfrac{|B^+_k(r)|^2 k^{1.6}}{B_0^2(r)}
\dfrac{\partial}{\partial k} \dfrac{|B^+_k(r)|^2k^{1.6}}{4\pi},
\label{+8}
\end{array}
\end{equation}
Equation for characteristics of~(\ref{+8}) take a form:
\begin{equation}
\dfrac{1}{k^{1.6}}\dfrac{d k}{d r}=
\kappa I\dfrac{[2\pi f_{br}(r)]^{0.4}V_A(r)^{0.6}}{u+V_A(r)},
\label{+9}
\end{equation}
where
$I=|B^+_{k_{br}}(r)|^2k_{br}/B_0^2(r) =5\cdot 10^{-4}$,
$u\gg V_A(r)$.
Beyond the $187.5R_\odot$ 
(0.87  astronomical units --- AU) the collisional hydromagnetic
turbulence spectrum of Kraichnan~\cite{Krai} with the spectral
index
$\alpha=-1.5$~\cite{Tu}
replace the collisionless spectra with the spectral indexes
$\alpha=-1$ and
$\alpha=-1.6$.

\begin{figure}
\vspace{120mm}
\caption{The characteristics of the equation (30).}
\end{figure}

Integrating~(\ref{+9}) and transforming the result to the
 frequency dependence on the radial distance we obtain:
\begin{equation}
f=\left[
C
\left(
     \dfrac{r}{R_\odot}
\right)^{0.6}-
\left(   \dfrac{1}{f_{br}(r)}
\right)^{0.6}
\right]^{-1.6},
\label{+10}
\end{equation}
where $C$ is an arbitrary constant.

Figure 1 shows the characteristics of the equation~(\ref{+8})
 between the curves of the spectral break frequency 
$f_{br}$
and the ion cyclotron frequency
$f_{ci}=1.6\cdot 10^4\left(R_\odot/r\right)^2$,
where the wave energy is totally absorbed in the ion cyclotron
 resonance.

To estimate the fraction of Alfv\'en wave energy transferred to
 the protons of solar wind plasma in the process of induced
 scattering of waves we use,
 following Tu~\cite{Tu1}, the equation for the magnetic moment
 of protons.
Doing this we take into account that due to the smallness of
 the left-hand side of the equation~(\ref{+8}) the product
$|B^+_k(r)|^2 k^{1.6}$
has a very weak dependence on the wave vector
$k$.
Therefore the right-hand side of~(\ref{+8}) also depends weakly on
$k$
and we can rewrite it in form:
\begin{equation}
\left[
      u(r)+V_A(r)
\right]
\dfrac{\partial}{\partial r} \dfrac{|B^+_k(r)|^2}{4\pi}
 = -\dfrac{\partial}{\partial k} 0.25 \kappa f_{br}(r) I^2
B_0^2(r)\equiv -\dfrac{\partial}{\partial k} \Pi(r),
\label{+11}
\end{equation}
where
$\Pi(r)$
is the volumetric energy flux.

The equation for the magnetic moment take a form:
\begin{equation}
u \dfrac{d}{dr}\ln
\left(
       \dfrac{T_\perp}{B}
\right) = 
\dfrac{\Pi(r)}{n(r)k_B T_\perp},
\label{+12}
\end{equation}
where
$ T_\perp $
is the perpendicular proton temperature in Kelvin, 
$k_B$
 Boltzmann's constant,
$ n(r)=10^8 (R_\odot/r)^2$ cm${}^{-3}$
density of plasma.

We integrate the equation~(\ref{+12}) from the three solar
 radii assuming that the acceleration of solar wind is
 already achieved.
As a result we have:
\begin{equation}
\dfrac{T_\perp}{B_0}-\dfrac{T_\perp}{B_0}
 = \int\limits_{3R_\odot}^r
\dfrac{\Pi(r)}{n(r)k_B u B_0(r)}dr =
0.144 \ln
\left[
     0.6 +\pi\kappa\ln\dfrac{r}{R_\odot}.
\right],
\label{+13}
\end{equation}

Non-adiabatic heating of the protons related to the presence of
 the volumetric energy flux
$\Pi(r)$
reach only 
$0.16$ K$/$nT
in the range of radial distances
$[3R_\odot,200R_\odot]$.
Therefore the source of energy for the acceleration of fast
 solar wind from the polar coronal holes must now be found in
 the solar corona itself~\cite{Marsch}.

Below the spectral brake frequency 
$f_{br}(r)$
we have the spectrum of
 Livshits-Tsytovich~\cite{LivTsy},~\cite{Tsytovich}
 with the spectral index 
$\alpha=-1$.
However this spectrum doesn't extend from
$f_{br}$
to the whole low frequency domain but truncated by the WKB
 solution.

Solving the stationary equation for Alfv\'en 
waves~\cite{Whang},~\cite{Holl},~\cite{Barnes}:
\begin{equation}
\nabla
\left(\left(
      3 {\bf u}(r)\pm 2{\bf V}_A(r)
\right)
\dfrac{|B^+_k(r)|^2k}{8\pi}
\right) - {\bf u}(r)\nabla
\dfrac{|B^+_k(r)|^2 k}{8\pi}=0,
\label{+14}
\end{equation}
the equation for the wave amplitude was found in the form:
\begin{equation}
\left[
      1\pm \dfrac{V_A(r)}{u}
\right]^2 n(r)^{-3/2}|B^+_k(r)|^2k
= {\rm const}.
\label{+15}
\end{equation}
At the distances
$r>10R_\odot$
we can neglect the small parameter
$V_A(r)/u\ll 1$
and rewrite it in the form defining the reduction of the wave
 amplitudes and their spectral index
$\alpha=-2/3$~\cite{Tu}:
\begin{equation}
{\rm const}\cdot n(r) =
\left(
    |B^+_k(r)|^2k
\right)^{2/3}.
\label{+16}
\end{equation}
Besides that, knowing from observations that the spectral
 break curve 
$f_{br}$
and the WKB solution are merging at the distance 0.87 AU 
(187.5 $R_\odot$)
we are able to find the form and position of WKB solution
 curve from the equation~(\ref{+16}) in the form (fig.~1):
\begin{equation}
f_{WKB}(r)= 1.6\cdot 10^{-3}\cdot 
\left(\dfrac{r}{187.5R_\odot}\right)^{2/3}\text{{} ľ}.
\label{+17}
\end{equation}

\section{Conclusion}

Using the  collisionless kinetic equation for 
 interacting Alfv\'en waves in the random phase approximation
 we have followed analytically the evolution of these waves in the
 process of their propagation from the polar solar corona.
As a result of such evolution the boundary of waves with the
 spectral index 
$\alpha=-1$
is shifted towards low frequencies due to induced scattering
 of waves by ions, which increase energy of ions by the recoil
 effect. 
However, this spectrum does not extend over the whole low
 frequency range but is truncated by WKB solution.

Above the frequency
$f_{br}$
the spectrum with the spectral index
$\alpha=-1.6$
is obtained.
Let us note that both in the polar solar corona and in the
 interplanetary space the approximation of collisionless
 plasma is valid at least up to the distance 0.87 AU.
\par\bigskip
A.~A.~Galeev thanks the Alexander von Humboldt Stiftung for
 support and Prof.~Joachim Tr\"umper for attention to this work.
This paper was partially supported by the RFBR grants
 97-02-16489 and 96-15-96723 (science school grant).


\end{document}